\documentclass[aps,floatfix,twocolumn,pra,nofootinbib,superscriptaddress,longbibliography]{revtex4-2}
\usepackage{bm,color}
\usepackage{array}
\usepackage{amsmath,amsfonts,amssymb,stackrel}
\usepackage{longtable}
\usepackage{float}
\usepackage[T1]{fontenc}
\usepackage[utf8]{inputenc}
\usepackage{bm}
\usepackage{ulem}
\usepackage{tikz}
\usetikzlibrary{calc,patterns,decorations.pathmorphing,decorations.markings}
\PassOptionsToPackage{hyphens}{url}

\newcommand{\be}{\begin{equation}}
\newcommand{\ee}{\end{equation}}
\newcommand{\bea}{\begin{eqnarray}}
\newcommand{\eea}{\end{eqnarray}}

\newcommand{\bb}[1]{\left( #1 \right)}
\newcommand{\bbr}[1]{\left. #1 \right)}
\newcommand{\bbl}[1]{\left( #1 \right.}
\newcommand{\bbcro}[1]{\left[ #1 \right]}
\newcommand{\bbcror}[1]{\left. #1 \right]}
\newcommand{\bbcrol}[1]{\left[ #1 \right.}

\newcommand{\ii}{\textrm{i}}
\newcommand{\dd}{\textrm{d}}
\newcommand{\eee}{\textrm{e}}
\newcommand{\rr}{\textbf{r}}
\newcommand{\qq}{\textbf{q}}
\newcommand{\kk}{\textbf{k}}

\DeclareMathOperator\ch{ch}

\newcommand{\hk}[1]{{ #1}}

\tikzstyle{ressort}=[decorate,decoration={zigzag,pre length=0.0cm,post length=0.0cm,segment length=5, amplitude=0.1cm}]

\tikzset{->-/.style={decoration={
  markings,
  mark=at position .5 with {\arrow[scale=2,color=black]{>}}},postaction={decorate}}}
\tikzset{-<-/.style={decoration={
  markings,
  mark=at position .5 with {\arrow[scale=3,color=black]{<}}},postaction={decorate}}}

\tikzset{->>-/.style={decoration={
  markings,
  mark=at position .5 with {\arrow{>>}}},postaction={decorate}}}
\tikzset{-<<-/.style={decoration={
  markings,
  mark=at position .5 with {\arrow{<<}}},postaction={decorate}}}

\tikzset{phantom->-/.style={decoration={
  markings,
  mark=at position .5 with {\arrow[scale=2]{>}}},postaction={decorate}}}

\tikzset{serpent/.style={decoration={snake},postaction={decorate}}}

\usepackage[breaklinks=true]{hyperref}
\hypersetup{
    colorlinks,
    linkcolor={red!50!black},
    citecolor={blue!90},
    urlcolor={blue!80!black}
}
\usepackage{cleveref}
\AtBeginEnvironment{thebibliography}{%
  \renewcommand{\emph}[1]{\uline{#1}}%
}

\begin{document}
\title{Plasmons in three-dimensional superconductors}

\author{T. Repplinger}
\affiliation{Laboratoire de Physique Théorique,
Université de Toulouse, CNRS, UPS, 31400, Toulouse, France}
\author{S. Klimin}
\affiliation{TQC, Universiteit Antwerpen, Universiteitsplein 1, B-2610 Antwerp, Belgium}
\author{M. Gélédan}
\affiliation{Laboratoire de Physique Théorique,
Université de Toulouse, CNRS, UPS, 31400, Toulouse, France}
\author{J. Tempere}
\affiliation{TQC, Universiteit Antwerpen, Universiteitsplein 1, B-2610 Antwerp, Belgium}
\author{H. Kurkjian}
\affiliation{Laboratoire de Physique Théorique,
Université de Toulouse, CNRS, UPS, 31400, Toulouse, France}

\begin{abstract}
We study the plasma branch of an homogeneous three-dimensional electron gas 
in an $s$-wave superconducting state. Although a sum rule guarantees that the 
departure of the plasma branch always coincides with the plasma frequency $\omega_p$, the 
dispersion and lifetime of plasmons is strongly affected by the presence of the 
pair condensate, especially when $\omega_p$ is close to the 
pair-breaking threshold $2\Delta$. When $\omega_p $ is between $1.7\Delta $ and 
$2\Delta$, the level repulsion is strong enough to give the plasma branch an anomalous, 
downward dispersion and a dispersion minimum strictly lower than $\omega_p$. Then 
for $\omega_p>2\Delta$, plasmons damp out in pair-breaking excitations, acquiring a 
small damping rate at zero temperature, which we compute in a non-perturbative way.
Finally, the density-density response function displays a resonance near $2\Delta$ (not to be confused with the amplitude mode), 
which can beat with the main plasma resonance, and subsists for $\omega_p$
large compared to $\Delta$, thereby distinguishing charged from neutral condensates.

\end{abstract}
\maketitle

\section{Introduction} Despite being a very mature experimental platform, supporting numerous technical 
applications, superconductors still hold some of the most fundamental open questions
of many-body physics. The impressively high critical temperature ($T_c$) and the unconventional
Cooper pairing in cuprates and iron-based superconductors are the most famous
of those fascinating questions. However, even some properties of conventional Bardeen-Cooper-Schrieffer 
(BCS) superconductors are still intensively discussed, such as the existence of an amplitude collective mode 
\cite{Sacuto2014,Measson2019,Shimano2013}, reminiscent of the Higgs mode in high-energy physics.

In fact, even for such usual behavior as plasma oscillations (the collective modes of the electronic density),
superconductors are still not fully understood. In a pioneering work, Anderson \cite{Anderson1958} has shown 
that the phononic (Goldstone) branch that exists in a neutral fermionic condensate 
acquires a gap corresponding to the plasma frequency $\omega_p$ in presence of long-range Coulomb interaction. 
This mechanism later became famous due to its analogy with the phenomenon of mass acquisition in high-energy physics.
The work of Anderson has then been revisited in the context of high-$T_c$ superconductivity 
\cite{DasSarma1991,Griffin1993,Kobelkov1995,vanderMarel1995,Takada1997plasm},
and nuclear/neutronic matter \cite{Ducoin2011}. While Anderson focused on the regime
of large $\omega_p$, the frequency of transverse plasmons in layered materials (such as cuprates) 
softens to an acoustic dispersion, such that in the superconducting phase an undamped plasma branch can be expected\cite{Uchida1992,Richard1994,Tachiki1997,KeJinZhou2020} below 
the pair-breaking threshold $2\Delta$. While a sum-rule\cite{Takada1997plasm} guarantees 
that the branch always departs from $\omega_p$, the dispersion relation
was shown\cite{DasSarma1995,DasSarma1995manip,Yamada1998} to approach the phononic law ($\omega_{q,{\rm n}}$) of neutral fermionic condensates 
as $\sqrt{\omega_p^2+\omega_{q,{\rm n}}^2}$ in the limit $\omega_p\ll2\Delta$.

In both limits of large and small $\omega_p$, the dispersion of plasmons
is thus similar to the dispersion in the normal phase, and no significant effect of superconductivity
has been reported so far. On the contrary, our study identifies a significant distorsion
of the plasma resonance caused by superconducting electrons. The distorsion is largest when
$\omega_p$ is close to $2\Delta$, but even for large $\omega_p$ the pair-breaking continuum
bears the trace of Coulomb interactions. 

We consider the reference situation of an isotropic three-dimensional 
(3D) $s$-wave superconductor but our study can be readily extended to layered geometries or anisotropic pairing.
We identify three main differences between normal and superconducting plasmons.
First, when $1.696\Delta<\omega_p<2\Delta$, the plasma branch is repelled by 
the pair-breaking threshold, and acquires an anomalous dispersion, with a negative 
curvature and thus a minimum strictly below $\omega_p$. Second, at $\omega_p>2\Delta$,
plasmons are damped (even at zero temperature) and decay by breaking Cooper pairs. 
Last but not least, we find a second resonance in the low-energy region of the pair-breaking 
continuum, separated from the main plasma peak.
This second peak is particularly intense when $\omega_p$ is near $2\Delta$ and leads to spectacular beatings
in the time evolution of a perturbation of the electronic density. The peak however subsists in the regime $\omega_p\gg\Delta$,
and is thus a fingerprint of fermionic condensates with long-range interactions.

By assessing the influence of superconductivity on density oscillations,
our study can guide practical use of plasmonics to probe and manipulate 
superconducting materials \cite{Cavalleri2013,Basov2014,KeJinZhou2020}.
The low-energy plasmons we describe may also affect the
critical temperature through their zero-point motion \cite{Pentegov2008}.

\section{Dispersion equation} 

We study an homogeneous electron gas evolving 
in a cubic volume $V$ with a average density $\rho$, defining the
Fermi wavenumber $\rho=k_F^3/3\pi^2$. Electrons interact through both the long-range Coulomb potential $V_C(r)\propto 1/r$ and 
a short-range part, responsible for $s$-wave Cooper pairing, and modelled by a contact potential
of coupling constant $g$:
\be
{V}(\rr_1,\rr_2)=g \delta(\rr_1-\rr_2)+V_{C}(\rr_1-\rr_2)
\ee
In terms of the electron 
mass $m$ and wavenumber $q$, the Fourier transform of the Coulomb potential 
is $V_C(q)=m\omega_p^2/\rho q^2$ (we use $\hbar=k_B=1$ throughout the article).

We imagine that the system is driven at fixed frequency $\omega$ and wavenumber 
$q$ by an external field (for example an electromagnetic field) and we study the 
collective response within linear response theory. 
In more standard situations this response can be described by London electrodynamics 
\cite{Buisson1997} (a long wavelength effective theory), but for the present purpose of describing 
the interplay between plasma waves and Cooper pairing, a microscopic theory, such as
the Random Phase Approximation (RPA) is unavoidable.
Such an approach results in a linear system \cite{repdens} relating the density $\delta\rho$ 
and pair-field fluctuations (in phase $\delta\theta$ and modulus $\delta|\Delta|$)
to the corresponding drive fields $u_\theta$, $u_{|\Delta|}$ and $u_\rho$:
\be
\begin{pmatrix} 2\ii\Delta\delta\theta(\qq,\omega) \\  2\delta|\Delta(\qq,\omega)| \\ 2V_C(q)\delta\rho(\qq,\omega) \end{pmatrix} = \chi(\omega,\qq)
\begin{pmatrix} u_\theta (\qq) \\ u_{|\Delta|}(\qq) \\  u_\rho(\qq)   \end{pmatrix},
\label{3par3}
\ee
The $3\times3$ response matrix $\chi$, which incarnates the coupling
between density and pairing fluctuations in superconductors,
is expressed (see Appendix \ref{chi}) in terms of the bare propagator $\Pi$
as $\chi=(D-\Pi)^{-1}\Pi$ with
\be
D\equiv\begin{pmatrix}
V/g &0&0\\0&V/g&0\\0&0&V/2V_C(\qq) 
\end{pmatrix}
\ee

The spectrum of the collective modes corresponds to the poles of $\chi$,
hence to the zeros of $M=\Pi-D$:
\be
\text{det}M_{\downarrow}(z_\qq,\qq)=0 \label{detM}
\ee
 When damping mechanisms are active (for example at $\omega>2\Delta$ or at non-zero temperature), a branch cut 
appears on the real axis, representing the coupling to the continuum of decay channels. In such situation, 
we use recently develop technics \cite{higgs,higgslong,artlongsk} to extract
the pole in an analytic continuation through the branch cut (in Eq.~\ref{detM},
$M_\downarrow$ denotes such an analytic continuation of $M$).
This study focuses on the typical 
weak-coupling regime of superconductors, with $\Delta$ much smaller 
than the Fermi energy $\epsilon_F$, and the excitation wavelength 
comparable to the Cooper pair size $\xi=k_F/2m\Delta$. In this regime,
the fluctuation of the modulus of the order parameter are decoupled from the
phase-density fluctuations:
\be
\text{det}M_{\downarrow}=0 \iff  M_{11,\downarrow}M_{33,\downarrow}-M_{13,\downarrow}^2=0 \text{ or } M_{22,\downarrow}=0\label{phasedensite}
\ee
The second condition gives rise the ``pair-breaking'' or ``Higgs'' modulus mode
which in the weak-coupling regime is insensitive to Coulomb interactions  \cite{Popov1976,higgslong}. 
Here, we study the density-phase modes, fulfilling the first condition.

\section{Anomalous dispersion of long wavelength plasmons}
\begin{figure}
\includegraphics[width=0.5\textwidth]{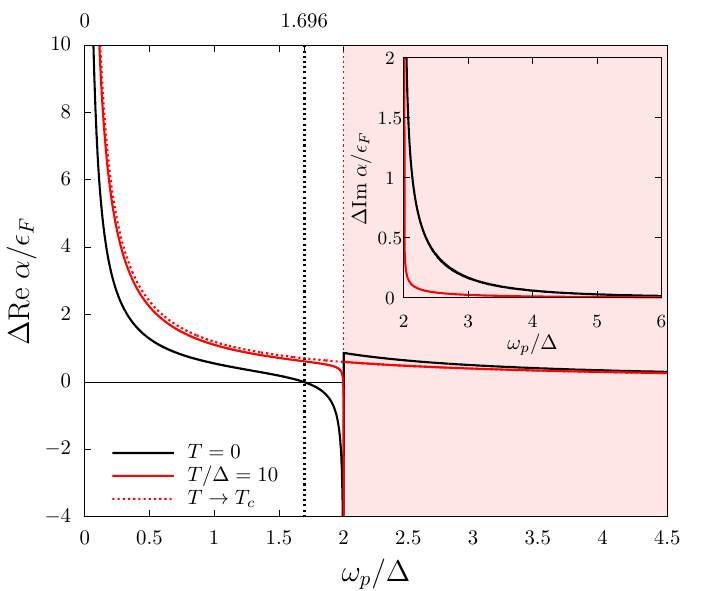}
\caption{\label{fig:alpha} Dispersion parameter $\text{Re\,}\alpha$ (multiplied by $\Delta/\epsilon_F$ to have a finite weak-coupling limit),
in function of the plasma frequency at zero (black curve) and high temperature (red curve). The normal dispersion $6\epsilon_F/5\omega_p$
is shown by the red dotted curve. The value where $\alpha$ changes sign at $T=0$ is indicated by the black dotted line.
Inset: the damping parameter $\text{Im\,}\alpha$, which becomes nonzero inside the pair-breaking continuum
$[2\Delta,+\infty[$ (red area).}
\end{figure}

We first study analytically the plasmon dispersion in the limit
$q\ll 1/\xi$, where, by analogy with the normal case \cite{FetterWalecka},
one can expect the quadratic law \cite{Kobelkov1995}:
\be
z_q=\omega_0+\alpha\frac{q^2}{2m}+O(q^4)
\label{zq}
\ee
A sum rule \cite{Takada1997plasm} guarantees that the origin $\omega_0$
of the plasma branch always coincides with the plasma frequency
\be
\omega_0=\omega_p
\label{omega0}
\ee
as in the normal phase. Superconductivity however greatly influences the departure of the
plasma branch through its curvature $\alpha$.
At zero temperature, we derive the fully analytic expression of $\alpha$:
\be
\alpha = \frac{6\epsilon_F}{5\omega_p}-\frac{32 \epsilon_F}{15\omega_p} \frac{\Delta^2 \text{arcsin}\bb{{ \omega_p}/{2\Delta}}}{\omega_p\sqrt{4\Delta^2-\omega_p^2}} \label{alpha}
\ee
which is shown as a black curve on Fig.~\ref{fig:alpha}. In the conventional limit 
$\omega_p\gg2\Delta$, we recover the normal plasmon dispersion \cite{FetterWalecka} $\alpha\to6\epsilon_F/5\omega_p$.
Highly-energetic density waves are thus insensitive to the weak-pairing between electrons.
In the opposite ``{quasiphononic}'' limit $\omega_p\ll2\Delta$, which 
corresponds to the experimental situation of Refs.~\cite{DasSarma1995manip,Tachiki1997},
rather than expanding for fixed $z$ as prescribed by \eqref{zq},
one should expand\cite{DasSarma1995} for $q\to0$ while keeping $z/v_F$ 
comparable to $q$. This yields\footnote{Note that this is consistent with
the behavior of $\alpha$ in the limit $\omega_p/\Delta\to0$.}
\be
z_\qq \underset{\substack{q\to0 \\ cq/\omega_p \text{ fixed}}}{\longrightarrow}{\sqrt{\omega_p^2+c^2q^2}}
\ee
where $c=v_F/\sqrt{3}$ is the speed-of-sound of the weakly-interacting condensate of neutral fermions.
This bending of the plasma branch to a linear dispersion when $\omega_p$ tends to 0 is similar
to what happens in the normal phase, where the normal plasma branch tends to zero-sound
(with the difference that the velocity of zero-sound is $v_F$ instead of $v_F/\sqrt{3}$ here).

The most remarkable behaviour occurs in between those two limits.
First, the repulsion of the pair-breaking threshold leads to a squareroot 
divergence of $\text{Re}\,\alpha$ when approaching the 
pair-breaking threshold from below. This opens an interval $\omega_p\in[1.696\Delta,2\Delta[$ where
plasmons have an anomalous \textit{negative} dispersion at the origin, that is, $\text{Re}\alpha<0$.
Then, at $\omega_p>2\Delta$, Eq.~\eqref{alpha} (with $\omega_p\to\omega_p+\ii0^+$) 
shows that $\alpha$ acquires an imaginary part that describes the nonzero damping rate of plasmons. 
This reflects the fact that a superconductor has pair-breaking decay channels
available even at zero-temperature (unlike the particle-hole channels of the normal phase).
Those channels are very active when $\omega_p$ is just above the pair-breaking threshold
such that $\text{Im}\alpha$ shows a squareroot divergence when $\omega_p\to 2\Delta^+$.
On the contrary, they weaken in the limit of large $\omega_p$, such that $\text{Im}\alpha$
vanishes as $-16\pi\Delta^2\epsilon_F/\omega_p^3$.

\hk{At nonzero temperature, plasmons are also sensitive to the quasiparticle-quasihole 
excitations.} Eq.~\eqref{alpha} generalizes into
\be
\alpha=\frac{\epsilon_F}{\Delta}\bbcro{\frac{6\bar\omega_p}{5}\bb{I_3+J_0-J_2}-\frac{8}{3\bar\omega_p}I_1}
\label{alphaTnonnulle}
\ee
in terms of the dimensionless parameters $\bar T=T/\Delta$, $\bar\omega_p=\omega_p/\Delta$
and the integrals $I_n=\int_{0}^{+\infty}\dd\xi\frac{\text{th}({\epsilon/2\bar T})}{\epsilon^n({\bar\omega_p^2-4\epsilon^2})}$
and $J_n=\frac{1}{2\bar T\bar\omega_p^2}\int_{0}^{+\infty}\frac{\dd\xi}{ \epsilon^n \ch^2({\epsilon/2\bar T})}$
with $\epsilon=\sqrt{\xi^2+1}$. The red curve in Fig.~\ref{fig:alpha} shows $\alpha$ 
in the vicinity of the critical temperature $T/T_c=0.9989$ ($T/\Delta=10$). We observe
that $\alpha$ tends to its normal limit $6\epsilon_F/5\omega_p$ uniformly
except in a neighborhood of size $\approx \Delta^2/T$ around $2\Delta$.
There, the divergence of the real and imaginary parts is preserved whenever $T<T_c$,
showing that a regime of anomalous plasmon dispersion subsists until the transition to the normal phase.
In usual situations, $\omega_p$ is fixed in units of the Fermi energy $\epsilon_F$, but
the ratio $\omega_p/\Delta(T)$ can still be adjusted by varying the temperature. The negative plasma dispersion
will thus eventually occur when increasing the temperature provided that $\omega_p$ is below $2\Delta$ at $T=0$.

One could be surprised that plasmons remain undamped ($\text{Im}\,\alpha=0$) for $\omega_p<2\Delta$
despite a nonzero temperature which provides a decay channel through quasiparticle-quasihole
excitations. In fact, to absorb a plasmon (\text{i.e.} to satisfy the resonance condition $\omega_p=\epsilon_{\qq+\kk/2}-\epsilon_{\qq-\kk/2}$)
quasiparticles need to have a wavenumber $k>2m\omega_p/q$. The plasmon lifetime thus follows an activation law
$\text{Im} z_q\propto \eee^{-2m\omega_p^2/q^2T}$ which is exponentially suppressed in the limit $\Delta/\epsilon_F,T/\epsilon_F\to0$
with $\omega_p,q$ of order $\Delta,1/\xi$. Plasmon damping at $\omega_p<2\Delta$ is thus essentially a strong-coupling
effect.

\section{Resonance splitting}
\begin{figure}
\includegraphics[width=0.5\textwidth]{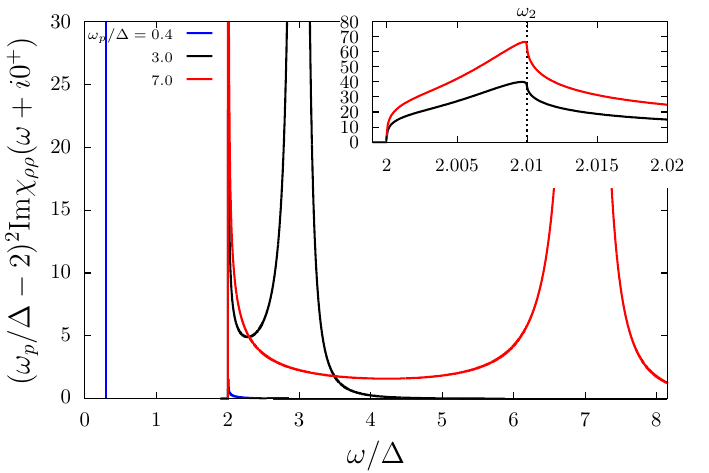}
\caption{\label{splitting}  Density-density response
in function of the excitation frequency $\omega$ at fixed wave number $q\xi=0.1$
and various plasma frequencies. For visibility, we have multiplied $\chi_{\rho\rho}$ by $(\omega_p/\Delta-2)^2$.The inset is a zoom on the first peak in the interval $[2\Delta,\omega_2]$.
On the blue curve, the Dirac peak is at $\omega\simeq0.416\Delta$.}
\end{figure}

Superconductivity not only bends the dispersion of the plasma branch, it also deforms
the shape of the density response function $\chi_{33}(\omega)$ at frequencies close to 
the pair-breaking threshold. Besides the Lorentzian peak centered around $\text{Re}z_\qq\simeq\omega_p$,
a second peak, shown on Fig.~\ref{splitting}, emerges between $2\Delta$ and the second branching point \cite{DasSarma1995,higgs}
\be
\omega_2=\sqrt{4\Delta^2+\epsilon_F \frac{q^2}{2m}}
\label{omega2}
\ee
of the continuum. The peak is absent in the neutral case (which in our case
corresponds to the limit $\omega_p\to 0$, see the blue curve in Fig.~\ref{splitting}), and starts to grow as $\omega_p$
approaches $2\Delta$ from below. It reaches its maximal intensity when
$\omega_p$ passes $2\Delta$ but the peak remarkably persists even
in the regime $\omega_p\gg 2\Delta$ (although its spectral weight relative
to the main plasma resonance decreases in this limit, red curve in Fig.~\ref{splitting}). 
It thus seems as if a part of the spectral weight
gets captured when $\omega_p$ passes the range $[2\Delta,\omega_2]$
and remains trapped in this range even when $\omega_p$ becomes large.
Similarly to the disparition of
the phononic branch, this peak above $2\Delta$ is thus a signature of long-range interactions,
with the difference that it is specific to the superconducting state (whereas the Anderson
mechanism occurs also in the normal phase).

The behavior of this peak becomes clearer when looking at the analytic structure of $\chi_{33}$. 
The Riemann sheet connected to the interval $[2\Delta,\omega_2]$ of the real axis
contains a unique pole of the density-phase propagator at
\begin{multline}
z_q^{\rm II}=2\Delta-\frac{\ii\text{sign}(2\Delta-\omega_p)}{\sqrt{\Delta}}\sqrt{\frac{8}{3\pi^2}\left\vert1-\frac{4\Delta^2}{\omega_p^2}\right\vert} \bb{\frac{k_Fq}{2m}}^{3/2}\\+O\bb{q^{7/4}}
\label{zII}
\end{multline}
This pole should not be confused with the famous amplitude ``Higgs'' mode \cite{Popov1976,higgs,Measson2019},
although both poles lie in the same energy range $[2\Delta,\omega_2]$, they concern excitation
channels (the density-phase channel for $z_q^{\rm II}$, the modulus channel for the amplitude mode)
which are decoupled in the weak coupling limit.
Eq.~\eqref{zII} exhibits an unusual non-integer power-law dispersion\footnote{
When $\omega_p=2\Delta$ the quadratic law reemerges $z_q^{\rm II}=2\Delta-(0.0184+0.9953\ii)\frac{\epsilon_F}{\Delta}\frac{q^2}{2m}+O(q^4)$.
This result, like Eq.~\eqref{zII}, are obtained by expanding at low $q$ as prescribed by Eq.~(10) in Ref.~\cite{higgs}.},
which contrasts with the quadratic dispersion of the plasma and amplitude modes.

The real part of $z_q^{\rm II}$ is either below $2\Delta$ when $\omega_p<2\Delta$ or above $\omega_2$
when $\omega_p>2\Delta$. 
This explains why the associated peak fades at low $\omega_p$ (and disappears in the neutral
case), and has its maximum in $\omega_2$ for $\omega_p>2\Delta$, as shown by the inset of Fig.~\ref{splitting}. 
Eq.~\eqref{zII} behaves well in the limit $\omega_p/\Delta\to+\infty$, which confirms that the peak near the continuum edge survives in this limit.

\begin{figure}
\includegraphics[width=0.5\textwidth]{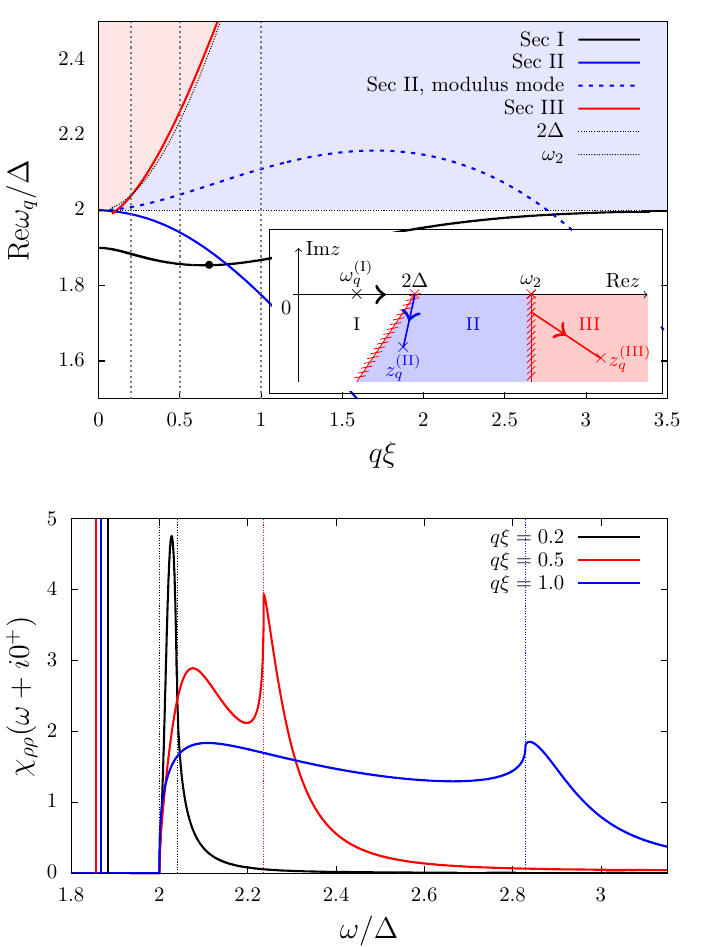}
\caption{\label{dispSeuil} (Top pannel) Eigenfrequency $\text{Re}z_q$ of the plasma branch in function
of the wave vector $q$ (in unit of the inverse pair radius $\xi=k_F/2m\Delta$),
with $\omega_p=1.9\Delta$. The angular
points $2\Delta$ and $\omega_2$ (Eq.~\eqref{omega2}) are shown as dotted lines.
The analytic windows are shown in colors: white for $\omega<\omega_1$ (window I), blue for $\omega_1<\omega<\omega_2$
(window II) and red for $\omega>\omega_2$ (window III). The solution of Eq.~\eqref{phasedensite} in
each window is shown as a solid line in the corresponding color. The inset shows their
schematic trajectories in the complex plane after analytic continuation. In window II, the pair-breaking
mode (solution of $M_{22,\downarrow}=0$) is shown as a dashed line. The dispersion minimum of the undamped
solution below $2\Delta$ is shown by the black dot.
(Bottom pannel) Density-density response
in function of the excitation frequency $\omega$ at fixed plasma frequency
$\omega_p=1.9\Delta$ and excitation wave number $q=0.2/\xi$, $0.5\xi$ and $1.0\xi$ (corresponding
to the vertical dotted lines in the top pannel).
The angular points $2\Delta$ and $\omega_2(q)$ are marked by vertical dotted
lines. Besides the Dirac peaks below $2\Delta$, broadened peaks are visible
inside the pair-breaking continuum.}
\end{figure}

Fig.~\ref{dispSeuil} summarizes the analytic structure of $\chi$. The function is divided in three analyticity windows
(I, II and III) by its two branching points $2\Delta$ and $\omega_2$. Each window is associated to a separate Riemann
sheet (inset of Fig.~\ref{dispSeuil}) each containing a single pole of the density-phase propagator (respectively $\omega_q^{\rm I}$, $z_q^{\rm II}$ and $z_q^{\rm III}$).
Conversely, $\chi_{22}$ only has a pole in sheet II, corresponding to the amplitude mode (dashed blue line).
Here, for $\omega_p=1.9\Delta$, the real pole $\omega_q^{\rm I}$ supports the main plasma branch departing in $\omega_p$
(while for $\omega_p>2\Delta$ the main branch would be supported by $z_q^{\rm III}$),
$z_q^{\rm II}$ is below $2\Delta$ and $z_q^{\rm III}$ follows rather closely the angular point $\omega_2$. 

\section{Beating of density waves}
In the frequency domain, we have describe an unusual splitting of the plasma resonance into 
a peak around $\omega_p$ and a peak in the range $[2\Delta,\omega_2]$. To further illustrate
the originality of this phenomenon, we study its counter-part in the time-domain,
through the relaxation of abrupt density perturbations. 
Namely, we suppose that at $t>0$ an operator suddenly turns on a static external field $u_\rho(\rr)=u_0\cos(\qq\cdot\rr)$
coupled to the electronic density. The subsequent evolution of the
density perturbation is given by the inverse Laplace transform of the density-density response function:
\be
\delta\rho(\rr,t)=\frac{u_0\cos(\qq\cdot\rr)}{2V_C(q)}  \int_{+\infty+\ii\eta}^{-\infty-\ii\eta}\frac{\dd z}{2\ii\pi}\frac{\eee^{-\ii z t}}{z}\chi_{33}(z,\qq)
\label{Laplace}
\ee

\begin{figure}
\includegraphics[width=0.49\textwidth]{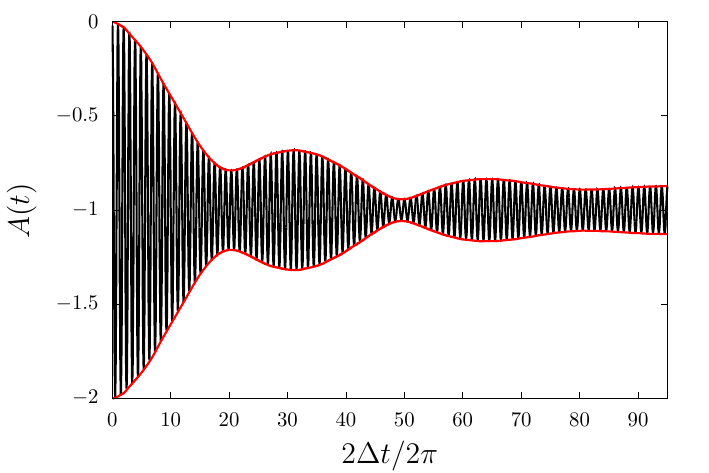}
\caption{\label{battements} Time-evolution of the amplitude $A(t)=2V_C(q)\delta\rho(\rr,t)/u_0\cos(\qq\cdot\rr)$ of 
a density wave created by a sudden excitation. We have used here $q\xi=0.1$
and $\omega_p=2.05\Delta$.}
\end{figure}

This integral can be closed into a winding contour around the branch cut $[2\Delta,+\infty[$ 
and a residue in the real pole $\omega_q^{\rm I}$.
The time-evolution of $\delta\rho$ thus combines the contributions of the plasmonic resonance
and of the peak near the continuum edge, which causes the system to oscillate
at multiple frequencies. As shown on Fig.~\ref{battements}, when the two peaks
are close (we use here $\omega_p=2.05\Delta$) this leads to very remarkable beatings, with a carrier oscillating at frequency
$\omega_p\approx 2\Delta$ modulated by an envelop of typical frequency $|\omega_p-2\Delta|\ll 2\Delta$.
As the contribution of the pair-breaking continuum to Eq.~\eqref{Laplace} decays with time,
the beatings gradually disappear and give way to undamped oscillations
at frequency $\omega_q^{\rm I}$ (very close to $2\Delta$ here) about the static
response $\delta\rho(\rr,t)=-\frac{u_0\cos(\qq\cdot\rr)}{2V_C(q)}$.

\section{Conclusion}
We have described the low-$q$
dispersion of superconducting plasmons in 3D, revealing a regime of anomalous downward
dispersion, and a finite lifetime due to the pair-breaking decay channels. 
A new resonance also emerges near the pair-breaking threshold, indicating
a splitting of density waves into high- and low-frequency components. 
For a more realistic description of plasmons in cuprates, 
our study should be extended to 2D superconductors \cite{Millis2020},
or 2D-layered electron gases \cite{Yamada1998,KeJinZhou2020,Wu2021}.
Our work may also be applied to superfluids of ultracold fermions \cite{Zwerger2012}
where different kind of long-range interactions can be engineered \cite{Ferlaino2018},
or neutron star matter \cite{Ducoin2011}.

\appendix

\section{Bare propagator}
\label{chi}

Here, we recall the expression of the bare propagator $\Pi$
which is used to construct the response function $\chi$.
The matrix elements can be expressed in a generic form
\begin{widetext}
\be
\Pi_{ij}(z,\qq)=\sum_\kk \frac{1}{2\epsilon_+\epsilon_-}\bbcro{ \frac{(1-f_+-f_-)\pi_{ij}^+}{z^2-(\epsilon_++\epsilon_-)^2} - \frac{(f_+-f_-)\pi_{ij}^-}{z^2-(\epsilon_+-\epsilon_-)^2}}
\label{piij}
\ee
with $\xi_\pm=(\qq/2\pm\kk)^2/2m-\mu$, $\epsilon_\pm=\sqrt{\xi_\pm^2+\Delta^2}$ and $f_\pm=1/(1+e^{\epsilon_\pm/T}$, and the (symmetric) matrices of coefficients
\bea
\pi^+&=& \begin{pmatrix}{(\epsilon_+ +\epsilon_- )\bb{\epsilon_+ \epsilon_- + \xi_+\xi_- + \Delta^2}} & z\bb{\epsilon_-  \xi_++ \epsilon_+  \xi_-} & - z(\epsilon_++\epsilon_-)\\ 
\ast & (\epsilon_+ +\epsilon_- )\bb{\epsilon_+ \epsilon_- + \xi_+\xi_- - \Delta^2} & -(\epsilon_+ +\epsilon_- )\bb{\xi_+ + \xi_-} \\
\ast & \ast & (\epsilon_+ +\epsilon_- )\bb{\epsilon_+ \epsilon_- - \xi_+\xi_- + \Delta^2} \end{pmatrix} \\
\pi^-&=&\begin{pmatrix} (\epsilon_+ -\epsilon_- )\bb{\epsilon_+ \epsilon_- - \xi_+\xi_- - \Delta^2}&z\bb{\epsilon_-  \xi_+- \epsilon_+  \xi_-} & z\bb{\epsilon_+-\epsilon_-}\\
\ast&(\epsilon_+ -\epsilon_- )\bb{\epsilon_+ \epsilon_- - \xi_+\xi_- + \Delta^2}&(\epsilon_+ -\epsilon_- )\bb{\xi_++\xi_-} 
\\ \ast &\ast& (\epsilon_+ -\epsilon_- )\bb{\epsilon_+ \epsilon_- + \xi_+\xi_- - \Delta^2} \end{pmatrix}
\eea
To compute the matrix $M=\Pi-D$ in long-wave limit, we perform a combinaison of lines and columns:
\be
N=\begin{pmatrix}  M_{11} &  M_{12} &   z  M_{13}+2 M_{11} \\  M_{12}& M_{22}&  z M_{23}/\Delta+2 M_{12}\\
   z M_{13}/\Delta+2 M_{11} &  z  M_{23}/\Delta+2 M_{12}&  z^2 M_{33}/\Delta^2+4  z  M_{13}/\Delta+4 M_{11}\end{pmatrix}
\label{Mtilde}
\ee
The advantage of this recombined matrix is that the whole third line and column is
of order $q^2$. Explicitely
\bea
N_{13}&=&\sum_\kk \frac{2\xi_+\xi_--\xi_+^2-\xi_-^2}{2\epsilon_+\epsilon_-}\bbcro{ \frac{(1-f_+-f_-)(\epsilon_++\epsilon_-)}{z^2-(\epsilon_++\epsilon_-)^2} + \frac{(f_+-f_-)(\epsilon_+-\epsilon_-)}{z^2-(\epsilon_+-\epsilon_-)^2}}\\
N_{23}&=&-\sum_\kk \frac{z(\xi_+-\xi_-)}{2\epsilon_+\epsilon_-}\bbcro{ \frac{(1-f_+-f_-)(\epsilon_+-\epsilon_-)}{z^2-(\epsilon_++\epsilon_-)^2} + \frac{(f_+-f_-)(\epsilon_++\epsilon_-)}{z^2-(\epsilon_+-\epsilon_-)^2}}\\
\Delta^2 N_{33}&=&
\frac{\rho L^3 q^2}{2m} \bb{1-\frac{z^2}{\omega_p^2}} \\
&&+\sum_\kk \frac{(\xi_+-\xi_-)^2 }{2\epsilon_+ \epsilon_-} \bbcro{
 \frac{ \bb{\epsilon_+ + \epsilon_-}\bb{1-f_+-f_-} (\epsilon_+\epsilon_- - \xi_+\xi_- -\Delta^2)}{ z^2-(\epsilon_++\epsilon_-)^2}
- \frac{ \bb{\epsilon_+ - \epsilon_-}\bb{f_+-f_-} (\epsilon_+\epsilon_- + \xi_+\xi_- +\Delta^2)}{ z^2-(\epsilon_+-\epsilon_-)^2}} \notag
\eea
Note that the expression of $N_{33}$ has been simplified using the sum rule found in Ref.~\cite{Takada1997plasm}, namely
\be
\sum_\kk\frac{
\bb{1-f_+-f_-}\bb{\epsilon_+ + \epsilon_-}\bb{\epsilon_+ \epsilon_--\xi_+ \xi_--\Delta^2}-\bb{f_+-f_-}\bb{\epsilon_+ - \epsilon_-}\bb{\epsilon_+ \epsilon_-+\xi_+ \xi_-+\Delta^2}}{2\epsilon_+\epsilon_-}
=\frac{\rho V q^2}{2m}.
\ee
Thus, to leading order in $q$, the eigenenergy of the plasma branch solves $N_{33}(z_q,q)=0$, which yields immediately Eq.~\eqref{omega0} of the main text.
\end{widetext}

{\section{Low-$q$ expansion of the fluctuation matrix}

We give additional detail on the low-$q$ expansion of $M$, which leads to expressions \eqref{alpha}
and \eqref{alphaTnonnulle} of the dispersion parameter in the main text. In this appendix, we use 
the dimensionless variables $\bar q=q\xi$, $\bar z=z/\Delta$ and 
\be
n_{ij}=N_{ij}\times{(2\pi)^3\epsilon_F}/{k_F^3 L^3} \label{mij}/4\pi
\ee
Generically, the expansion of a matrix element can be written as
\be
n_{ij}= \sum_{p=0}^n n_{ij}^{(p)}(\bar z)q^{2p}+O(q^{2(n+1)}) \label{devM}
\ee
and all coefficients $n_{ij}^{(p)}$ are elementary functions of $\bar z$, of the function
\bea
g(\bar z)=&
=\begin{cases}-\frac{\text{arcsin}\bb{{ \bar z}/{2}}}{\bar z\sqrt{4-\bar z^2}} \quad\text{if}\quad \bar z<2 \\\frac{\text{argcosh}\bb{{\bar z}/{2}}}{\bar z\sqrt{\bar z^2-4}} -\frac{\ii\pi}{2\bar z\sqrt{\bar z^2-4}}\quad\text{if}\quad \bar z>2  \end{cases} \label{g}
\eea
and its derivative. Their explicit expression is given in Table~\ref{tablegn}.
\begin{table}[htb]
\begin{center}
\begin{tabular}{|c|c|c|}
\hline
Coefficient &At $T=0$ & At $T\neq0$ \\
 \hline &&\\
$n_{33}^{(0)}$ & $0$&$0$\\
$n_{33}^{(2)}$ & $\frac{2}{3}\bb{1-\frac{ \bar z^2}{ \bar \omega_p^2}}$ & $\frac{2}{3}\bb{1-\frac{ \bar z^2}{ \bar \omega_p^2}}$ \\
$n_{33}^{(4)}$ &  $\frac{8 (4 g+1)}{5 \bar z^2}$ & $\frac{8}{5}(I_3-J_2+J_0)$ \\
$n_{33}^{(6)}$ & $\frac{32 \left(-8 \bar z g'+8 \left(\bar z^2-1\right) g+\bar z^2-2\right)}{7 \bar z^6}$  & \\ && \\
 \hline && \\
$n_{11}^{(0)}$ & $\frac{ \bar z^2 g}{2} $ & $\frac{ \bar z^2 I_1}{2} $ \\
$n_{11}^{(2)}$ &$-\frac{4 \bar z {g}'+4 {g}+1}{3 \bar z^2} $ & \\ && \\
 \hline && \\
$n_{13}^{(0)}$ & $0$ & $0$ \\
$n_{13}^{(2)}$ & $-\frac{4g}{3} $ & $-\frac{4}{3}I_1 $\\
$n_{13}^{(4)}$ &$\frac{8 \left(4 \bar z g'-2 \left(\bar z^2-2\right) g+1\right)}{5 \bar z^4}$ & \\ && \\
 \hline
\end{tabular}
\end{center}
\caption{\label{tablegn} Long-wavelength expansion of the elements of $N$ in terms of the function $g$ of Eq.~\eqref{g} (at $T=0$) and
of the integrals $I_n$ and $J_n$ defined below Eq.~\eqref{alphaTnonnulle} (at $T\neq0$).}
\end{table}

At $T=0$ we have derived the coefficient of the term in $q^4$
in the plasma dispersion (such that $z_\qq=\omega_0+\alpha\frac{q^2}{2m}+\frac{\beta}{\epsilon_F}\bb{\frac{q^2}{2m}}^2+O(q^6)$):
\be
\beta = \frac{\epsilon_F^3}{\omega_p^3} h\bb{\frac{\omega_p}{\Delta}} \label{beta}
\ee
with
\begin{multline}
h(\omega)= \frac{1}{1575 \omega^2}\bbcrol{64 \omega g'(\omega) \left(112 \omega^2 g(\omega)+63 \omega^2-220\right) }\\ \bbcror{-10 \left(32 g(\omega) \left(56 \omega^2
   g(\omega)-9 \omega^2+44\right)+27 \omega^2+352\right)}
\end{multline}
This coefficient $\beta$ is positive in the interval $[1.696\Delta,2\Delta]$ where $\alpha$ is negative,
which allows us to estimate the position of the dispersion minimum as 
\be
q_{\rm min}\xi\approx\sqrt{-\frac{\epsilon_F^2}{\Delta^2}\frac{\alpha}{2\beta}}
\label{approxqmin}
\ee
Fig.~\ref{minplasm} show the dependence of $q_{\rm min}$ on the ratio
$\omega_p/\Delta$, using both \eqref{approxqmin} and the exact numerical solution.
\begin{figure}[htb]
\includegraphics[width=0.5\textwidth]{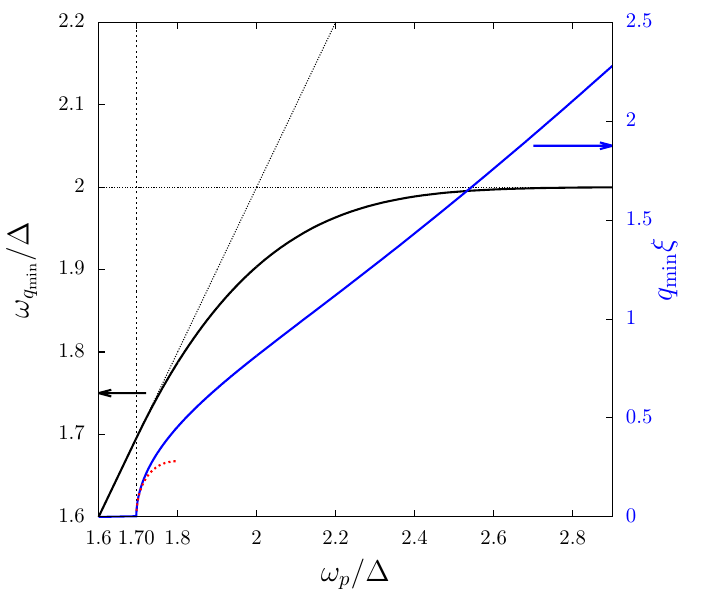}
\caption{\label{minplasm} The dispersion minimum $\omega_{q_{\rm min}}$ and the 
wavenumber $q_{\rm min}$ at which it is reached in function of the plasma frequency. For $\omega_p<1.696$,
$q_{\rm min}$ is identically 0 and $\omega_{q_{\rm min}}$ coincides with $\omega_p$ (oblique dotted line).
Then for $\omega_p>1.696$, $q_{\rm min}$ departs from 0 as described by Eq.~\eqref{approxqmin} (dashed red curve).
In the limit $\omega_p\to+\infty$, $q_{\rm min}$ diverges linearly and $\omega_{q_{\rm min}}$ tends to $2\Delta$.}
\end{figure}

\section{Numerical implementation}

Here we give additional details on how to evaluate the matrix $M$ numerically at $T=0$ but
arbitrary values of $q\xi$. The associated Fortran code is available online \cite{codeArxiv220111421}.
The rigorous way to take the BCS limit ($\Delta\to0$) and to deal with 
the resonance condition is explained in Ref.~\cite{higgslong}. 
For $\Delta\to0$ and fixed $\bar q=q\xi$, $\bar\omega=\omega/\Delta$, the momentum integrals 
defining the matrix element $M_{ij}$ are dominated by wavevectors close to the dispersion
minimum $k_0=\sqrt{2m\mu}\simeq k_F$. We thus set $\bar\xi=\xi_k/\Delta$,
$ k\dd k/2m\Delta=\dd\xi/2$, and expand the integrand for $k_0\gg \sqrt{2m\Delta}$.

The integral over $\bar\xi$ from $-\infty$ to $+\infty$ is odd in the case of $M_{12}$ and $M_{23}$ (which therefore vanish),
and even otherwise. The spectral density associated to $m_{ij}$ (the dimensionless version of $M_{ij}$, see Eq.~\eqref{mij})
takes the form
\be
\rho_{ij}(\bar\omega)=\int_0^{+\infty} \dd \bar\xi \int_{0}^1 {\dd u} \bar\pi_{ij}^+\delta(\bar \omega- \bar\epsilon_+ -\bar\epsilon_-)
\label{rhoij}
\ee
with
\bea
\bar\pi_{11}^{+}&=&  \bar\epsilon_+\bar \epsilon_- +\bar \xi_+\bar\xi_- + 1 \\
\bar\pi_{22}^{+} &=&\bar\epsilon_+ \bar\epsilon_- + \bar\xi_+\bar\xi_- - 1 \\
\bar\pi_{33}^{+} &=&\bar\epsilon_+\bar \epsilon_- - \bar\xi_+\bar\xi_- + 1 \\
\bar\pi_{13}^{+} &=& - \bar\epsilon_+-\bar\epsilon_- 
\eea
We use here (and everywhere in this appendix) the dimensionless notations 
$\bar\xi_\pm=\xi_{\qq/2\pm\kk}/\Delta=\bar\xi\pm u\bar q$ and $\bar\epsilon_{\pm}=\sqrt{\bar\xi_\pm^2+1}$.

We use the Dirac delta to integrate over $\bar\xi$ at fixed $u$. The resonance condition ($\omega= \epsilon_+ + \epsilon_-$) is studied in Annexe A.
of Ref.~\cite{higgslong}. On the interval $[0,+\infty[$, it yields a unique root:
\begin{multline}
\bar\xi_0=\frac{\bar\omega}{2}\frac{r_2}{r_1}\text{ with } r_1={\sqrt{\bar\omega^2-4\bar q^2u^2}} \\ \text{ and } r_2={\sqrt{\bar \omega^2-4\bar q^2u^2-4}}
\end{multline}
$\xi_0$ is real provided
\be
u<u_{\rm max}=\bb{\frac{\bar\omega^2-4}{4\bar q^2}}^{1/2}
\ee
such that the remaining interval of integration over $u$ is
\be
I_u(\omega)=\begin{cases}
\emptyset \text{ if } \bar\omega<2 \\
[0,u_{\rm max}]\text{ if } 2<\bar\omega<\bar\omega_2\\
[0,1]                 \text{ if } \bar\omega>\bar\omega_2
\end{cases}
\ee
where $\bar\omega_2=2\sqrt{1+\bar q^2}$ is the dimensionless version
of Eq.~\eqref{omega2}.
After integration over $\xi$, we obtain
\be
\rho_{ij}= \int_{I_{u}(\omega)}  r_{ij}(u){\dd u}
\label{rhoij2}
\ee
with the integrands $r_{ij}$:
\bea
r_{11} &=& \frac{r_1}{4r_2} \\
r_{22} &=& \frac{r_2}{4r_1} \\
r_{33} &=& \frac{\bar \omega^2}{r_1^3 r_2} \\
r_{13} &=& \frac{\bar \omega}{2r_1 r_2} 
\eea
The angular integrals \eqref{rhoij2} can be computed analytically in terms of elliptic integrals
(denoted in the convention of Ref.~\cite{Gradshteyn}).
For $2<\bar\omega<\bar\omega_2$, we set $m=\sqrt{\bar\omega^2-4}/\bar\omega$, and obtain:
\bea
\rho_{11}&=&\rho_{33}=\frac{\bar\omega}{8\bar q}E(m)\\
\rho_{13}&=&\frac{K(m)}{4 \bar q}\\
\rho_{22}&=&\rho_{11}-\frac{2}{\bar\omega}\rho_{13}
\eea
For $\bar \omega>\bar \omega_2$, we perform the change of variable $\sin\phi=\sqrt{\bar \omega^2-4}u/2\bar q$.
Introducing $\theta=\text{arcsin}(2 \bar q/\sqrt{\bar\omega^2-4})$, we have
\bea
\rho_{11}&=&\frac{\bar \omega}{8\bar q}E(\theta,m)\\
\rho_{13}&=&\frac{F(\theta,m)}{4 \bar q}\\
\rho_{22}&=&\rho_{11}-\frac{2}{\bar \omega}\rho_{13}\\
\rho_{33}&=&\rho_{11}-\sqrt{\frac{\bar \omega^2-4\bar q^2-4}{\bar \omega^2-4\bar q^2}}
\eea

Once the spectral densities are known, the value of $m$ at arbitrary $z$ is given
by frequency integrals
\bea
\!\!\!\!\!\!\!\!\!\!\!\!m_{ii}(\bar z) \!\!\!&=&\!\!\!\!\int_{2}^{+\infty}\!\!\!\!\dd\bar\omega\bbl{\rho_{ii}(\omega)\bbcro{\frac{1}{\bar z-\bar \omega}-\frac{1}{\bar z+\bar \omega}}} \notag \\&& \qquad \qquad \bbr{+\frac{1}{2\sqrt{\bar \omega^2-4}}} \text{ for }i=1,2 \label{mii} \\
\!\!\!\!\!\!\!\!\!\!\!\!m_{33}(\bar z) \!\!\!&=&\!\!\!\!\int_{2}^{+\infty}\!\!\!\!\dd\bar\omega\bbcro{\rho_{33}(\bar \omega)\bbcro{\frac{1}{\bar z-\bar\omega}-\frac{1}{\bar z+\bar\omega}}} -\frac{2\bar q^2}{3\bar \omega_p^2}  \label{m33}  \\
\!\!\!\!\!\!\!\!\!\!\!\!m_{13}(\bar z) \!\!\!&=&\!\!\!\!\int_{2}^{+\infty}\!\!\!\!\dd\bar\omega\bbcro{\rho_{13}(\bar \omega)\bbcro{\frac{1}{\bar z-\bar\omega}+\frac{1}{\bar z+\bar\omega}}}  \label{m13}
\eea
where we have used the trick of Ref.~\cite{higgslong} to handle the regularizing counter-term $-V/g$: 
we subtract $M_{11}(\omega=0,\qq)=0$ to $M_{11}(\omega,\qq)$ and $M_{22}(\omega=2\Delta,\qq)=0$ to $M_{22}(\omega,\qq)$
and we use the expression of the spectral densities at zero wave vector 
$\rho_{22}(\bar\omega,0)={\sqrt{\bar\omega^2-4}}/{4\bar\omega}$ and $\rho_{11}(\bar\omega,0)=\bar\omega/4{\sqrt{\bar\omega^2-4}}$.
We have used also the (im)parity of the spectral densities $\rho_{ii}(-\bar\omega)=\rho_{ii}(\bar\omega)$
and $\rho_{13}(-\bar\omega)=-\rho_{13}(\bar\omega)$.

Note that the integral form [\eqref{mii}--\eqref{m13}] remain valid in the vicinity of the real axis ($\bar z=\bar \omega_0+\ii 0^+$), in
which case they should be understood as principal parts.
To deal with the cancellation of the denominator, we write
\begin{multline}
\mathcal{P}\int_{\omega_1}^{\omega_2}\dd\omega{\frac{\rho_{ij}(\omega)}{\omega_0-\omega}}=\int_{\omega_1}^{\omega_2}\dd\omega{\frac{\rho_{ij}(\omega)-\rho_{ij}(\omega_0)}{\omega_0-\omega}}\\-\rho_{ij}(\omega_0)\log\left\vert\frac{\omega_0-\omega_2}{\omega_0-\omega_1}\right\vert
\label{Piij}
\end{multline}
To reach a good precision on the integrand, one should be careful to split it at its angular point $\bar\omega_2$.
A change of variable may also be needed to handle the $1/\sqrt{\bar\omega-2}$ divergence at the continuum edge.

\paragraph*{Analytic continuation}
To analytically continue $M$ through window II or III, we use the formula of Nozières:
\be
m_{\downarrow}^{(\text{II or III})}(z,\qq)=
\begin{cases}m(\bar  z,\bar q) \text{, }\text{Im}\,z>0\\
m(\bar z,\bar  q)-2\ii\pi\rho^{(\text{II or III})}_{\downarrow}(\bar z,\bar q) \text{, }\text{Im}\,z<0
\end{cases}
\ee
where $\rho^{(\text{II or III})}_{\downarrow}$ is the analytic continuation of the spectral density from
the interval $[2,\bar\omega_2]$ or $[\bar\omega_2,+\infty[$ onto the lower-half complex plane. 
In practice, it is much easier to analytically continue $\rho$ than $M$ directly, which is why the formula of Nozières
is useful. In the present case, it is enough to complexify the integral expression
\eqref{rhoij2}. This means that $(i)$ the integrand becomes complex (in particular because the resonance energy $\bar\xi_0$
becomes complex), $(ii)$ the integration interval $I_u$ can become a contour in the complex plane. 
This contour can be deformed to optimize the convergence of the integral (as long as one stays away from
the branching points of the integrand).}

\begin{acknowledgments}

We acknowledge financial support from the Research Foundation-Flanders (FWO-Vlaanderen) Grant No. G.0618.20.N, and from the research council of the University of
Antwerp.

\end{acknowledgments}

\bibliography{/Users/hkurkjian/Documents/latex/HKLatex/biblio}

\end{document}